# Universal Phase Transitions of $B$1-Structured Stoichiometric Transition Metal Carbides

Zhisheng Zhao,[1] Xiang-Feng Zhou,[2] Li-Min Wang,[1] Bo Xu,[1] Julong He,[1] Zhongyuan Liu,[1] Hui-Tian Wang,[2] and Yongjun Tian [1,*]

[1]*State Key Laboratory of Metastable Materials Science and Technology, Yanshan University, Qinhuangdao 066004, China* ;

[2] *School of Physics and Key Laboratory of Weak-Light Nonlinear Photonics, Ministry of Education, Nankai University, Tianjin 300071, China*

The high-pressure phase transitions of $B$1-structured stoichiometric transition metal carbides (TMCs, TM=Ti, Zr, Hf, V, Nb, and Ta) were systematically investigated using *ab initio* calculations. These carbides underwent universal phase transitions along two novel phase-transition routes, namely, $B$1→distorted TlI (TlI′)→TlI and/or $B$1→distorted TiB (TiB′)→TiB, when subjected to pressures. The two routes can coexist possibly because of the tiny enthalpy differences between the new phases under corresponding pressures. Four new phases result from atomic slips of the $B$1-structured parent phases under pressure. After completely releasing the pressure, taking TiC as a representative of TMCs, only its new TlI′-type phase is mechanically and dynamically stable, and may be recovered.

---

[*] To whom correspondence should be addressed. Electronic addresses: fhcl@ysu.edu.cn

# 1. Introduction

$B$1 (NaCl-type)-structured compounds are very common. Examples include alkali halides, alkaline-earth metal oxides, lead chalcogenides, and transition metal light-element compounds. Their high-pressure behavior is the most intensively studied subject in high-pressure physics/chemistry, material science, and geoscience. Most $B$1-structured compounds, such as alkali halides and alkaline-earth metal oxides, follow the $B1 \rightarrow B2$ (CsCl-type) phase-transition route under pressure,[1–5] and several transition mechanisms have been used to interpret this phase transition.[6–12] Some transition metal oxides take the $B1 \rightarrow B8$ (NiAs-type) and $B1 \rightarrow R$ (rhombohedral structure) phase-transition routes.[13–16] However, to date, studies on the phase transition of $B$1-structured transition metal carbides (TMCs) are rare and seem to have yielded no significant results. Dubrovinskaia *et al.* reported that a $B1 \rightarrow R$ phase transition of TiC occurred at a pressure above 18 GPa.[17] Winkler *et al.* reevaluated these experiments and found no phase transition for TiC at pressures up to 26 GPa.[18] Liermann *et al.* found no phase transitions of $VC_{0.85}$ until 53 GPa; of NbC until 57 GPa; and of $TaC_{0.98}$ until 76 GPa, respectively.[19–21] These mean that the stabilities of $B$1-structured TMCs seem unshakable, and the high-pressure phases of TMCs are yet to be determined.

In the present paper, the high-pressure phase transitions of $B$1-structured stoichiometric TMCs (TM=Ti, Zr, Hf, V, Nb, and Ta) were systematically investigated using *ab initio* calculations. All investigated TMCs underwent two novel phase-transition routes, namely, $B1 \rightarrow$ distorted TlI (TlI′) $\rightarrow$ TlI and/or $B1 \rightarrow$ distorted TiB (TiB′) $\rightarrow$ TiB, which are highly likely to coexist because of the tiny enthalpy differences between the new phases under corresponding pressures. However, the thermodynamically favorable pathway is $B1 \rightarrow$ TlI′ $\rightarrow$ TlI, which results from the slightly lower enthalpies of the TlI′-type phases compared with the TiB′-type phases. Moreover, the stabilities of the new phases were further studied.

The emphasis points of the present study are as follows: (1) the TlI′- and TiB′-type phases, which were previously never found as the high-pressure phase in $B$1-structured compounds, are the new high-pressure phases in TMCs; (2) the TlI- and TiB-type phases were experimentally found as the intermediate states in the $B1 \rightarrow B2$ phase transition of AgCl and PbTe, respectively.[22, 23] However, they are first found as the final high-pressure phases for TMCs. No evidence indicates that the conventionally considered $B1 \rightarrow B2$ phase transitions predicted by Ahuja occur;[24] (3) the TlI′-type phase has a slightly lower enthalpy than the TiB′-type phase under relative pressures. However, for the TlI- and TiB-type phases, identifying which has the lower enthalpy is difficult because of the difference in the two calculation methods; (4) beyond the phase-transition pressures of TlI′→TlI and TiB′→TiB, the TlI′- and TiB′-type phases cannot retain their structures and thus change to the TlI- and TiB-type phases after full optimizations, respectively, and vice versa; and (5) at the ground state, taking TiC as a representative example of TMCs, only its new TlI′-type phase is mechanically and dynamically stable. The other three phases (TlI-, TiB′-, and TiB-type phases) are dynamically stable under high pressure.

## 2. Computation Methods

Calculations were performed using the CASTEP,[25] VASP,[26] USPEX,[27, 28] and CALYPSO codes.[29] All calculations were considered under the effect of pressure at 0 K. To maintain consistency, only the results obtained via the CASTEP code are shown in the figures and tables included in the current article.

Two methods were used to determine the high-pressure phases. The first involved "searching existing structures." The 14 existing structures investigated were as follows: the cubic structures $B$1 (SG, $Fm$-$3m$), $B$2 (SG, $Pm$-$3m$), zinc blende (SG, $Fd$-$3ms$), and FeSi (SG, $P$213); the orthorhombic structures TlI (SG, *Cmcm*), TiB (SG, *Pnma*), and MnP (SG, *Pnma*); the hexagonal structures WC (SG, *P*-6*m*2), NiAs (SG, *P*63/*mmc*), wurtzite (SG, *P*63*mc*), MoC (SG, *P*63/*mmc*), and TaN (SG, *P*-62*m*); the monoclinic structure RhSi (SG, *P*2$_1$/*c*); and the rhombohedral structure (*R* phase) FeO (SG, *R*-3*m*), which is the proposed structure of the TiC high-pressure phase in the controversial experiments.[17,18]

The calculations for the structural optimizations, enthalpies, and elastic stiffness constants were implemented in the CASTEP code based on the density functional theory (DFT). The Vanderbilt ultrasoft pseudopotential[30] was used, and the exchange-correlation function was treated by the Perdew-Berke-Ernzerhof (PBE) form of the generalized gradient approximation (GGA).[31] The cutoff energy for the plane wave basis set was chosen based on the criterion of ultrafine precision. A *k*-point separation (0.04 Å$^{-1}$) corresponding to the fine-quality level was used to generate the *k*-point grid resulting from the Monkhorst-Pack grid parameters.[32] Structural optimization was performed until the energy change per atom was less than $5 \times 10^{-6}$ eV, the forces on the atoms were less than 0.01 eV/Å, and all stress components were less than 0.02 GPa. An electronic smearing of 0.2 eV with a Gaussian scheme was employed.[33] This method does not reproduce the sharp features of the density of states (DOS), such as the van-Hove singularities, but produces a satisfactory general shape of the DOS even when a small number of *k*-points is used. In addition, the enthalpies under different pressures were recalculated using the GGA of the VASP code to ensure accurate calculations, and the phonon frequencies were calculated using the direct supercell method, which uses the forces obtained from the Hellmann-Feynman theorem.

The second method used the recently developed evolutionary and particle-swarm optimization algorithms for crystal structure prediction, and the most stable structures were predicted using the USPEX[27,28] and CALYPSO codes[29]. The details of the search algorithms and their applications were described in previous reports.[34–44] The underlying structure relaxations were performed using DFT within GGA as implemented in the VASP code.

## 3. Results and Discussion

△H-P Diagram. After full optimizations, only six structured TiC ($B1$, $B2$, zinc blende, WC, NiAs, and MoC structures) retained their initial crystal structures; the others were either distorted or changed to other structures under the given pressure ranges. The proposed $B1 \rightarrow R$ phase transition[17] will not occur because the high-energy $R$ phase changed directly to the $B1$ phase under pressure.

Figure 1 shows that four structured TiC always have higher enthalpies than the $B1$ phase under the considered pressures, and three high-pressure phase-transition routes were found. The $B1 \rightarrow TlI' \rightarrow TlI$ (Route 1) and $B1 \rightarrow TiB' \rightarrow TiB$ (Route 2) phase transitions occur under lower pressures compared with the $B1 \rightarrow B2$ (Route 3) phase transition. Therefore, the $B1 \rightarrow B2$ phase transitions for TiC proposed in Ref. 24 will not occur. The $B1 \rightarrow TlI'$ and $TlI' \rightarrow TlI$ transition pressures are 117 and 172 GPa, respectively, and those of $B1 \rightarrow TiB'$ and $TiB' \rightarrow TiB$ are 131 and 148 GPa, respectively. The two intermediate TlI'- and TiB'-type phases notably appeared. When full optimizations were performed, the TlI- and TiB-type phases directly changed to the TlI'- and TiB'-type phases in the 117–172 and 131–148 GPa pressure ranges, respectively. When the pressures are beyond 172 and 148 GPa, the TlI'- and TiB'-type phases transform into the TlI- and TiB-type phases, respectively. The two phase transition pathways are both driven by the lower enthalpies of the new phases.

The enthalpy differences between the TlI'- and TiB'-type phases and between the TlI- and TiB-type phases of TiC under pressures were calculated using the CASTEP and VASP codes, respectively (Table 1). The two codes yielded consistent results, indicating that the TlI'-type phase has a slightly lower enthalpy than the TiB'-type phase under relative pressures. For the TlI- and TiB-type phases, results obtained from the VASP code show that the former has slightly lower enthalpy than the latter. However, identifying which has the lower enthalpy using the CASTEP code is difficult because they nearly have the same enthalpies. On the other hand, the most stable structures under different pressures were found using the USPEX and CALYPSO codes. At 100, 120, 150, and 200 GPa, the most stable phases of TiC are the $B$1, TlI′-, TlI′-, and TlI-type phases, respectively. Furthermore, the TiB′- and TiB-type phases with slightly higher enthalpies were also found in our considered pressures. Therefore, the thermodynamically favorable pathway is $B1\rightarrow$TlI′$\rightarrow$TlI. However, the $B1\rightarrow$TiB′$\rightarrow$TiB route is likely to coexist with the $B1\rightarrow$TlI′$\rightarrow$TlI pathway because of the close enthalpies between the new phases under corresponding pressures.

Figure 2 shows the calculated enthalpy differences between the new phases and the $B$1 phases of the six TMCs compounds under different pressures along two different routes. All presented TMCs follow similar phase transition rules; the transition pressure points can be distinguished in Figure 2. After full optimizations, all TlI′-type phases of TMCs kept their structures at the ground state. Only the TiB′-type phases of TiC and VC kept their structures at the ground state; those of other TMCs changed into the $B$1 phases under lower pressures.

**P-V Diagrams.** The equations of states (P-V) of the TMCs along the two transition routes were determined. The volumes of the TlI′- and TiB′-type phases are almost similar under relative pressures, as well as the volumes of TlI- and TiB-type phases. For TMCs, both the $B1\rightarrow$TlI′ and $B1\rightarrow$TiB′ phase transitions exhibit an approximately 6% volume reduction, and the TlI′$\rightarrow$TlI and TiB′$\rightarrow$TiB phase transitions exhibit an approximately 0.5% volume reduction. Therefore, these phase transitions are first-order. The equations of states (P-V) of TiC, as a representative example of TMCs, are shown in Figure 3.

**Atomic Coordinate Numbers.** Figure 4 shows the top views of the TMCs high-pressure phase structures. The TM atoms have the same coordination numbers in the monoclinic TlI′-type (or TiB′-type) and orthorhombic TlI-type (or TiB-type) phases. However, the C atom has lower coordination numbers in the monoclinic phases because of the appearance of the nonbonding C-C. The coordination polyhedra for the TM and C atoms of the orthorhombic TlI- and TiB-type phases are plotted in Figure 5. High-pressure phase transitions result in increased atomic coordination numbers. The coordination numbers of the C atom are 6 in the initial $B$1 phase, 8 in the intermediate TlI′- and TiB′-type phases, and 9 in the final TlI- and TiB-type phases. For the metal atoms, their coordination numbers are from 6 in the initial $B$1 phase to 7 in the new phases.

**Bond Lengths.** For the TlI′- and TlI-type phases of TiC under 172 GPa (*i.e.*, the TlI′→TlI transition pressure), the TlI′-type phase shows lattice parameters of β=85.9°, a=7.246 Å, b=2.737 Å, and c=2.758 Å, compared with those of the TlI-type phases (caption of Figure 4). In the TlI′-type phase, every Ti atom is bonded with the surrounding 7 C atoms, and 4 types of Ti-C bonds with bond lengths of 2.058 Å (1 bond), 2.045 Å (2 bonds), 2.100 Å (2 bonds), and 1.987 Å (2 bonds) are found. Except for the 7 bonding Ti atoms, every C atom is still bonded to the nearest C atom, with a C-C bond length of 1.452 Å, and has a distance of 1.985 Å from the second nearest C atom. In the TlI-type phase, the Ti atom has the same 7 coordination numbers, and 3 types of Ti-C bonds, with bond lengths of 2.109 Å (1 bond), 2.081 Å (2 bonds), and 2.026 Å (4 bonds), exist. Compared with the coordination condition of the C atom in the TlI′-type phase, every C atom in the TlI-type phase is bonded to two C atoms, forming two similar C-C bonds with a bond length of 1.583 Å. By comparison, only one type of TM-C bond exists in the $B$1-structured TMCs, and the bonds are orthogonal to each other. For TiC, each Ti-C bond length is 1.933 Å under 172 GPa. Consequently, the lower volume (*i.e.*, higher density) (Figure 3) of the new phases results from the formation of short C-C bonds.

**Lattice Angle Changes.** Based on Figure 4, the TlI′- and TiB′-type phases can be regarded as small deformations of the TlI- and TiB-type phases, respectively. The β angles of the TlI′- and TiB′-type phases are continuously variable with increasing pressures until they reach the final TlI- and TiB-type states. Their β-angle changes are shown in Figure 6. The positions of anomalous mutations correspond to the phase-transition points, at which the enthalpies of the TlI′-type (TiB′-type) and TlI-type (TiB-type) are equal.

**DOS.** The color-contour maps of the charge densities and DOS of the TiC polymorphs at 140 GPa, as a representative of TMCs, are shown in Figure 7. The new phases have a total DOS at the Fermi level ($E_F$) that is nearly the same as that of the $B$1 phase. However, the contributions of the Ti 3$d$ and C 2$p$ electrons are different at $E_F$. In the $B$1 phase, because of the symmetric distribution of atoms and charges, the Ti 3$d$ and C 2$p$ electrons give almost identical contributions at $E_F$. In the new phases, strong C-C bonds and weak Ti-Ti couplings can be seen in the color-contour maps of the charge densities compared with the long-distance nonbonding Ti-Ti and C-C bonds in the $B$1 phase. The $d$ orbit has a three-dimensional petal-shaped electron cloud; thus, the $d$-$d$ orbit hybridizations of the metal atoms increase the electron DOS at $E_F$. By contrast, the strong $p$-$p$ orbit hybridizations of the short covalent C-C bonds improve the electronic tight-binding character and give decreasing contributions at $E_F$. From the $s$-$p$-$d$ hybridization to the $s$-$p$-$d$, $d$-$d$, and $p$-$p$ hybridizations, the pseudogap near $E_F$ in the $B$1 phase disappears. Instead, lowest DOS points in the new phases appear, resulting in increased stability.

**Phase-Transition Mechanisms.** A similar high-pressure structural rearrangement occurs in TMCs, and understanding the phase-transition mechanisms has become very important. The mechanisms for both routes involve atomic slips under pressure. In the $B1{\rightarrow}TlI'{\rightarrow}TlI$ and $B1{\rightarrow}TiB'{\rightarrow}TiB$ transitions, the atoms in the adjacent [001] directional layers of the $B1$ phase first undergoes an antiparallel displacement (Figure 8). Partial atoms then display unity slips of about a bond-length-unit distance along the [110] and [001] directions of the $B1$ phase in the $B1{\rightarrow}TlI'{\rightarrow}TlI$ and $B1{\rightarrow}TiB'{\rightarrow}TiB$ transitions, respectively. The $B1{\rightarrow}TlI'{\rightarrow}TlI$ transition is achieved through the aforementioned two-step change. However, the $B1{\rightarrow}TiB'{\rightarrow}TiB$ transition requires a third atomic slip of about a bond-length-unit distance along the [110] direction of the $B1$ phase.

**Stabilities of New High-pressure Phases.** To verify the mechanical stability of the high-pressure phases of TMCs, the elastic stiffness constants of the phases that kept their structures at the ground state after full optimizations were calculated (Table S1 in the Supporting Information). The criteria used to determine the mechanical stability of a monoclinic crystal are as follows:[45] $C_{11}>0$, $C_{22}>0$, $C_{33}>0$, $C_{44}>0$, $C_{55}>0$, $C_{66}>0$, $[C_{11}+C_{22}+C_{33}+2(C_{12}+C_{13}+C_{23})]>0$, $(C_{33}C_{55}-C_{35}^2)>0$, $(C_{44}C_{66}-C_{46}^2)>0$, $(C_{22}+C_{33}-2C_{23})>0$, $[C_{22}(C_{33}C_{55}-C_{35}^2)+2C_{23}C_{25}C_{35}-C_{23}^2C_{55}-C_{25}^2C_{33}]>0$, $\{2[C_{15}C_{25}(C_{33}C_{12}-C_{13}C_{23})+C_{15}C_{35}(C_{22}C_{13}-C_{12}C_{23})+C_{25}C_{35}(C_{11}C_{23}-C_{12}C_{13})]-[C_{15}^2(C_{22}C_{33}-C_{23}^2)+C_{25}^2(C_{11}C_{33}-C_{13}^2)+C_{35}^2(C_{11}C_{22}-C_{12}^2)+C_{55}(C_{11}C_{22}C_{33}-C_{11}C_{23}^2-C_{22}C_{13}^2-C_{33}C_{12}^2+2C_{12}C_{13}C_{23})]\}>0$. According to the criterion, the phases considered are all mechanically stable at the ground state. In addition, the bulk and shear moduli of the six $B1$-structured TMCs compounds were also calculated at ambient pressure (Table S2 in the Supporting Information). The simulated values were compared with available experimental data.

The dynamical stabilities of the new phases were also evaluated. Only the phonon frequencies of TiC in the considered TMCs compounds were calculated because of the tediousness of the computation. The TlI'-type phase at 0 GPa, TlI-type phase at 200 GPa, TiB'-type phase at 50 GPa, and TiB-type phase at 150 GPa have no imaginary phonon frequency within the whole Brillouin zone (Figure S1 in the Supporting Information). However, the TiB′-type phase at 0 GPa was dynamically unstable. These results indicate that the TiB′-type phase of TiC is stable under high pressure, and the TlI′-type phase may be recovered at ambient pressure. The ground-state equilibrium lattice parameters of the TlI′-type phases of TMCs were calculated in view of their possible recovery; the results are listed in Table 2.

Aside from temperature and time, pressure is another pivotal factor that determines the states of materials. Many materials exhibit novel structures and properties under pressure, which are mutative with the change in pressure. Studies in the field of high-pressure science offer significant contributions to geophysics and condensed material fields. Many have endeavored to increase the pressure limit of static laboratory experiments. In 1978, the pressure in the experiments with the diamond-window pressure cell exceeded 170 GPa.[46] In 1986, a diamond-anvil, high-pressure apparatus was used to extend the upper pressure limit from 210 to 550 GPa,[47] which is beyond the maximum pressure (360 GPa) of the earth's core. Therefore, the simulations in the present study can be validated in high-pressure experiments.

## 4. Conclusion

An overall report on the high-pressure phase transitions of $B1$-structured TMCs is presented. Details on the changes in the enthalpy, volume, atomic coordination numbers, bond lengths, lattice angles from the intermediate phases to the final phases, charge density and DOS, and stabilities of the new phases during the phase-transition process are included. Unlike in the phase transitions of other $B1$-structured compounds, two novel transition routes, namely, $B1\rightarrow$TlI′$\rightarrow$TlI and $B1\rightarrow$TiB′$\rightarrow$TiB, can coexist in TMCs because of the close enthalpies of the new phases under relative pressures. The intermediate TlI′- and TiB′-type phases are small deformations of the final TlI- and TiB-type phases, respectively. Along the two routes, the $B1\rightarrow$TlI′ (TiB′) and TlI′ (TiB′) $\rightarrow$TlI (TiB) phase transitions show volume reductions of approximately 6% and 0.5%, respectively. The new phases with high atomic coordination numbers have the lowest DOS values at $E_F$, indicating higher stabilities than the $B$1 phase under high pressure. The phase transitions result from the atomic slips of the parent $B$1 phase. Further research on the mechanical and dynamical stabilities of the new TiC phases show that the TiB′-type phase is stable under high pressure and that the TlI′-type phase may be recovered at ambient pressure.

**Acknowledgment.** This work was supported by NSFC (Grant Nos. 50872118, 50821001, 91022029, and 11174152), FANEDD (Grant No. 2007B36), NBRPC (Grant No. 2010CB731605 and 2011CB808205), Natural Science Foundation of Heibei Province of China (Grant No. E2009000453), and by the Science Foundation of Yanshan University for the Excellent Ph.D. Students (Grant No. YSUSF201101).

**Supporting Information Available:** The calculated phonon dispersion curves for the TiC (TlI′-type) phase at 0 GPa, the TiC (TlI-type) phase at 200 GPa, the TiC (TiB′-type) phase at 50 GPa, and the TiC (TiB-type) phase at 150 GPa are shown in Figure S1. The calculated elastic constants, bulk moduli $B$, and shear moduli $G$ of the $B$1- and high-pressure phases of TMCs at the ground state are listed in Tables S1 and S2. This information is available free of charge via the Internet at http://pubs.acs.org.

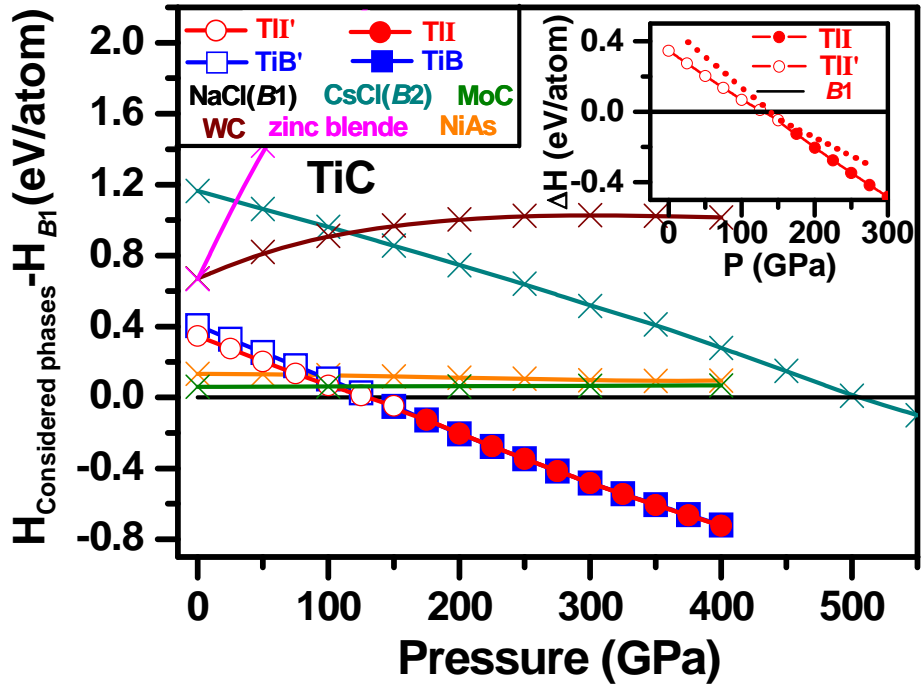

**Figure 1.** Computed enthalpy differences ($H_{\text{Considered phases}}-H_{B1}$) for TiC under different pressures. The inset shows the local region of the △H-P diagram along the $B1\rightarrow$TlI′$\rightarrow$TlI route. The dashed lines show that when full optimizations are conducted, the TlI′-type phase cannot retain its structure and directly transforms into the TlI-type phase beyond the TlI′→TlI phase-transition pressures, and vice versa.

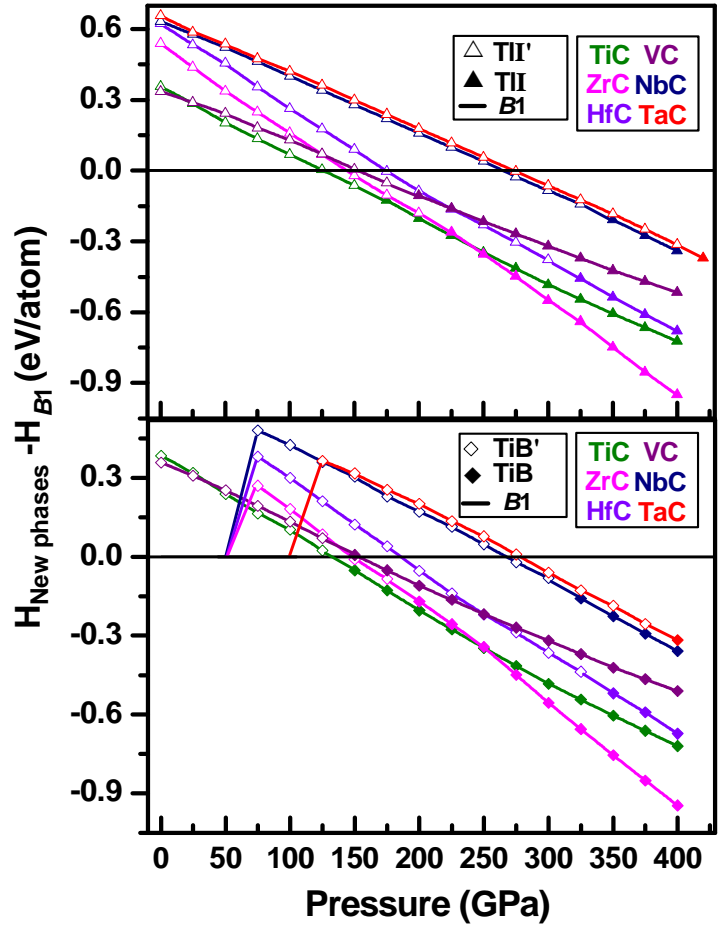

**Figure 2.** Computed enthalpy differences ($H_{\text{New phases}}-H_{B1}$) for TMCs along the $B1{\rightarrow}\text{TlI}'{\rightarrow}\text{TlI}$ (a) and $B1{\rightarrow}\text{TiB}'{\rightarrow}\text{TiB}$ (b) phase-transition routes.

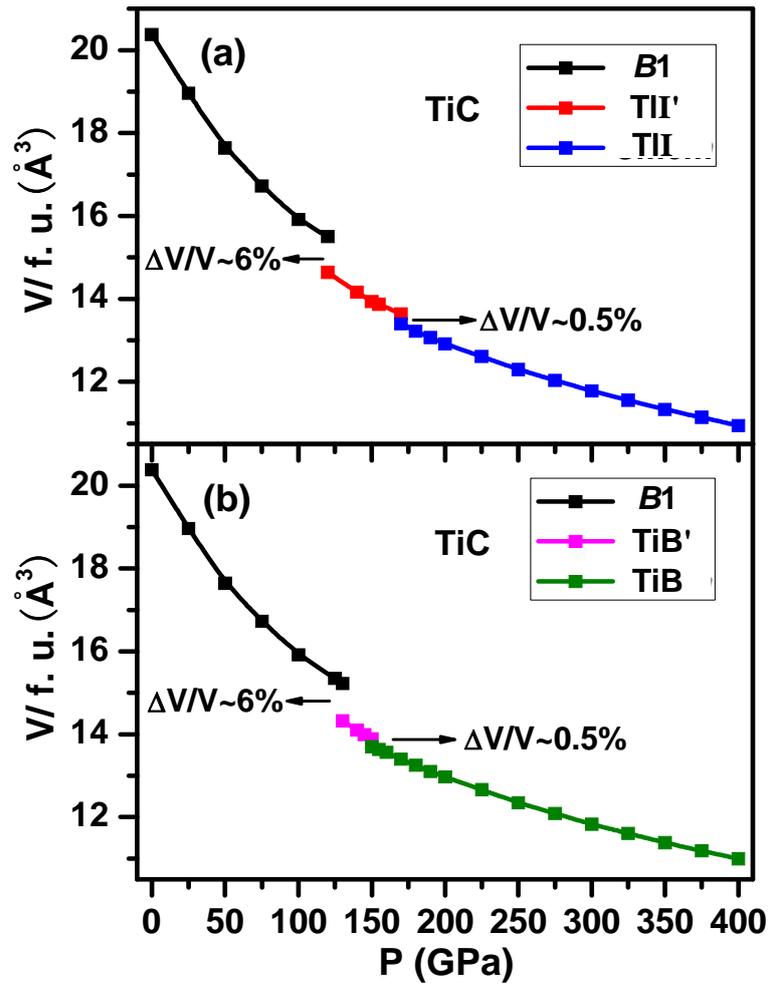

**Figure 3.** Equations of states (P-V) of TiC, as a representative example of TMCs investigated, along the two phase-transition routes.

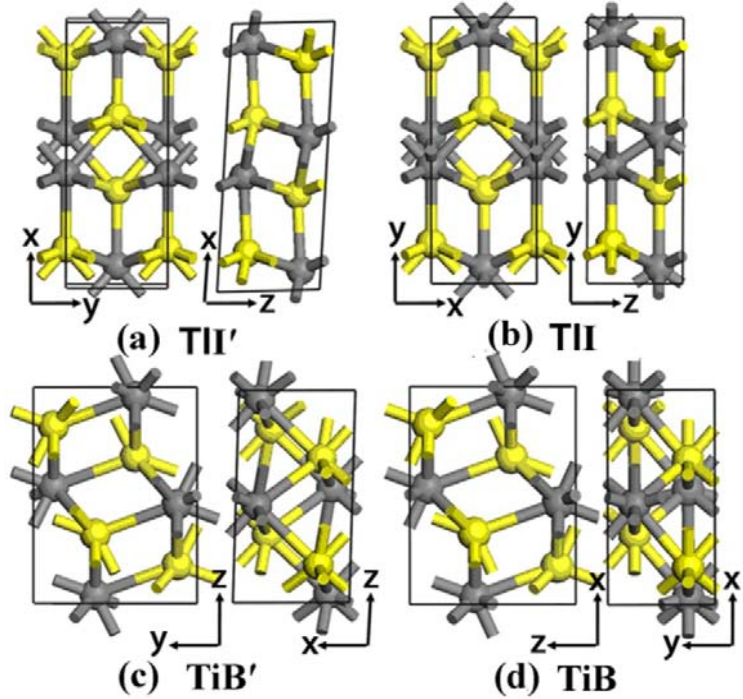

**Figure 4.** Top views of TMCs high-pressure phase structures. (a) and (c): the intermediate TlI′- and TiB′-type structures; (b) and (d): the final TlI- and TiB-type structures. The yellow and gray spheres represent metal atoms and carbon atoms, respectively. For TiC, TlI′-type phase at 117 GPa: space group ($C2/m$), α=γ=90°, β=84.9°, a=7.408 Å, b=2.802 Å, c=2.833 Å, Ti atoms occupying a 4$i$ position (0.856, 0, 0.728) and carbon atoms occupying a 4$i$ position (0.427, 0, 0.191); TlI-type phase at 172 GPa: space group ($Cmcm$), α=β=γ=90°, a=2.835 Å, b=7.265 Å, c=2.611 Å, Ti atoms occupying a 4$c$ position (0, 0.148, 0.75) and carbon atoms occupying a 4$c$ position (0, 0.439, 0.75); TiB′-type phase at 131 GPa: space group ($P2_1/c$), α=γ=90°, β=90.85°, a=2.748 Å, b=3.960 Å, c=5.260 Å, Ti atoms occupying a 4$e$ position (0.725, 0.620, 0.324) and carbon atoms occupying a 4$e$ position (0.802, 0.121, 0.461); TiB-type phase at 148 GPa: space group ($Pnma$), α=β=γ=90°, a=5.299 Å, b=2.625 Å, c=3.953 Å, Ti atoms occupying a 4$c$ position (0.327, 0.75, 0.612) and carbon atoms occupying a 4$c$ position (0.466, 0.75, 0.104).

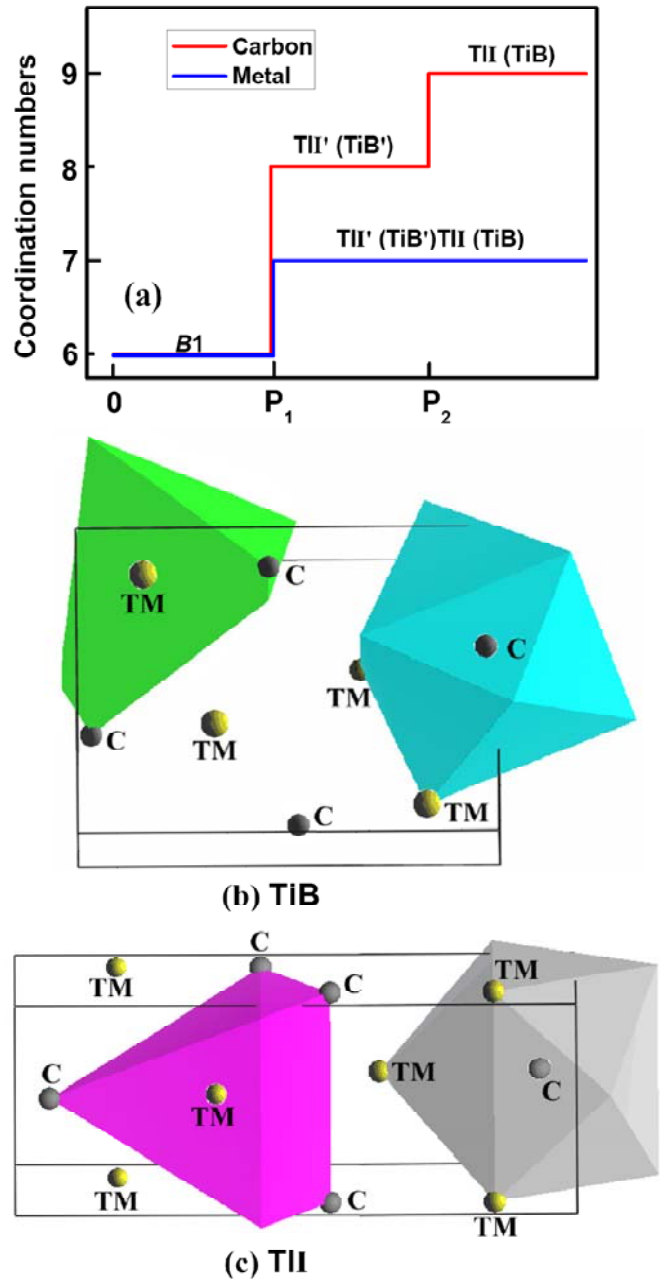

**Figure 5.** The coordination number changes (a) with phase transitions, and the coordination polyhedra for TM and C atoms of the orthorhombic TiB-type (b) and TlI-type (c) phases. The green and magenta polyhedra are for TM atoms; the cyan and gray polyhedra are for C atoms. $P_1$ and $P_2$ represent the indicative transition pressures of the $B1 \rightarrow$ monoclinic phases and monoclinic phases $\rightarrow$ orthorhombic phases, respectively.

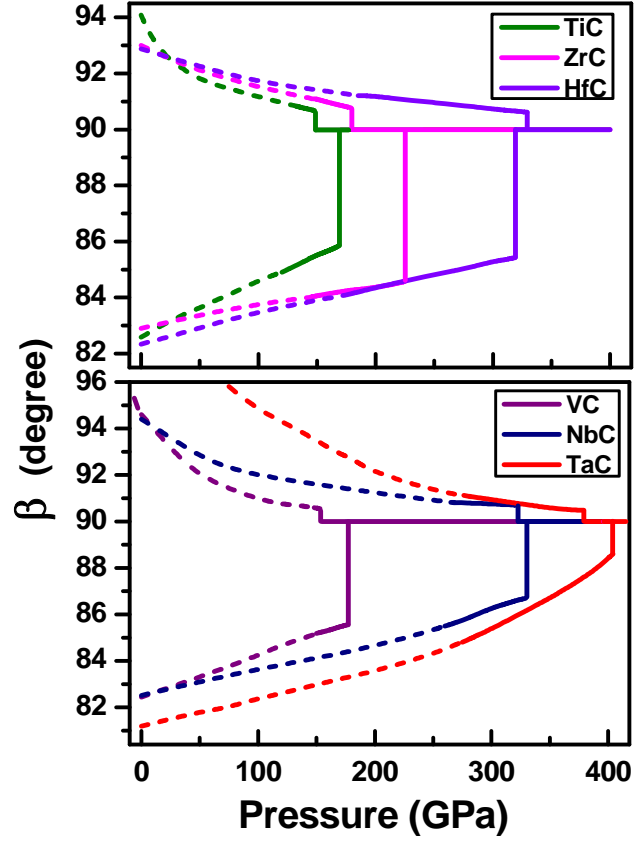

**Figure 6.** Lattice β angle changes from the TlI′- to TlI-type phases and TiB′- to TiB-type phases under pressures. For the TlI′→TlI transition, the β angle changes from <90° to 90° with increasing pressure. For the TiB′→TiB transition, the β angle changes from >90° to 90° with increasing pressure. The dashed lines represent the β angle changes of the TlI′-type (and TiB′-type) phase before the $B$1→TlI′ (TiB′) transition pressure.

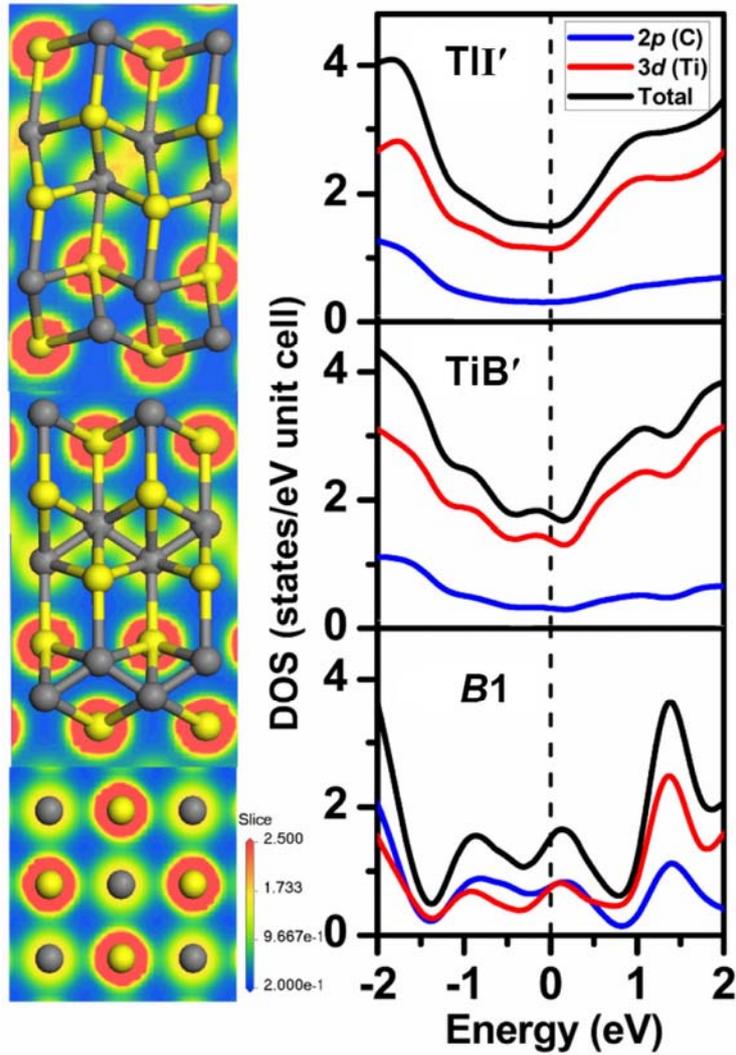

**Figure 7.** Color-contour maps of the total charge densities (left) and DOS (right) of the TiC polymorphs at 140 GPa. The tree maps of the charge densities are shown from top to bottom, corresponding to the (100) plane of the TlI′-type phase, the (904) plane of the TiB′-type phase, and the (100) plane of the $B$1 phase, respectively. The yellow and gray spheres represent the metal and carbon atoms, respectively.

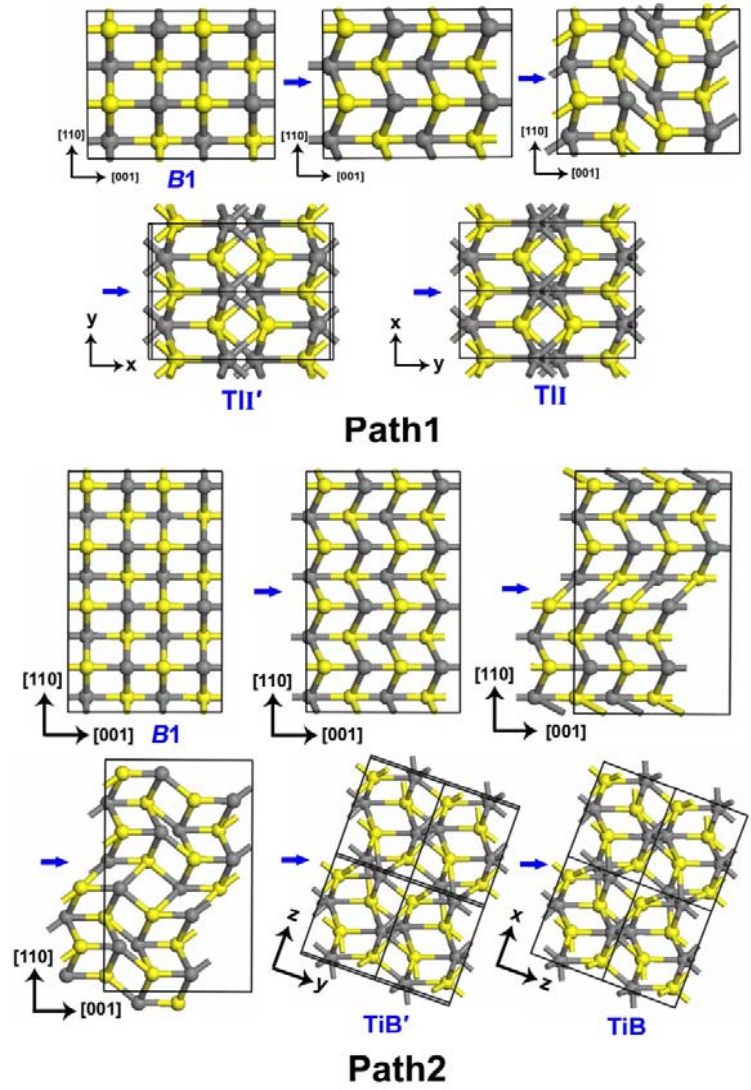

**Figure 8.** The visual renderings of TMCs phase transitions.

**Table 1.** Enthalpy differences ($H_{TII'}$-$H_{TiB'}$ and $H_{TII}$-$H_{TiB}$) for TiC under different pressures calculated using the CASTEP and VASP codes, respectively.

|  | CASTEP results | VASP results |  | CASTEP results | VASP results |
|---|---|---|---|---|---|
| P (GPa) | $H_{TII'}$-$H_{TiB'}$ (eV/atom) | $H_{TII'}$-$H_{TiB'}$ (eV/atom) | P (GPa) | $H_{TII}$-$H_{TiB}$ (eV/atom) | $H_{TII}$-$H_{TiB}$ (eV/atom) |
| 0 | -0.057 | -0.072 | 200 | 0.0023 | -0.019 |
| 25 | -0.056 | -0.069 | 250 | 0.0005 | -0.023 |
| 50 | -0.051 | -0.065 | 300 | 0.0001 | -0.024 |
| 75 | -0.045 | -0.059 | 350 | -0.0029 | -0.025 |
| 100 | -0.036 | -0.052 | 400 | -0.0038 | -0.022 |
| 120 | -0.029 | -0.046 |  |  |  |
| 140 | -0.018 | -0.041 |  |  |  |

**Table 2**. Ground-state equilibrium lattice parameters for the high-pressure TlI′-type phases of TMCs.

| TMCs | $a$ (Å) | $b$ (Å) | $c$ (Å) | $\beta$ (°) |
|---|---|---|---|---|
| TiC | 8.087 | 3.061 | 3.168 | 82.60 |
| ZrC | 8.842 | 3.324 | 3.451 | 83.01 |
| HfC | 8.950 | 3.315 | 3.402 | 82.30 |
| VC | 7.734 | 2.966 | 3.083 | 82.43 |
| NbC | 8.499 | 3.163 | 3.289 | 82.42 |
| TaC | 8.615 | 3.168 | 3.292 | 81.02 |

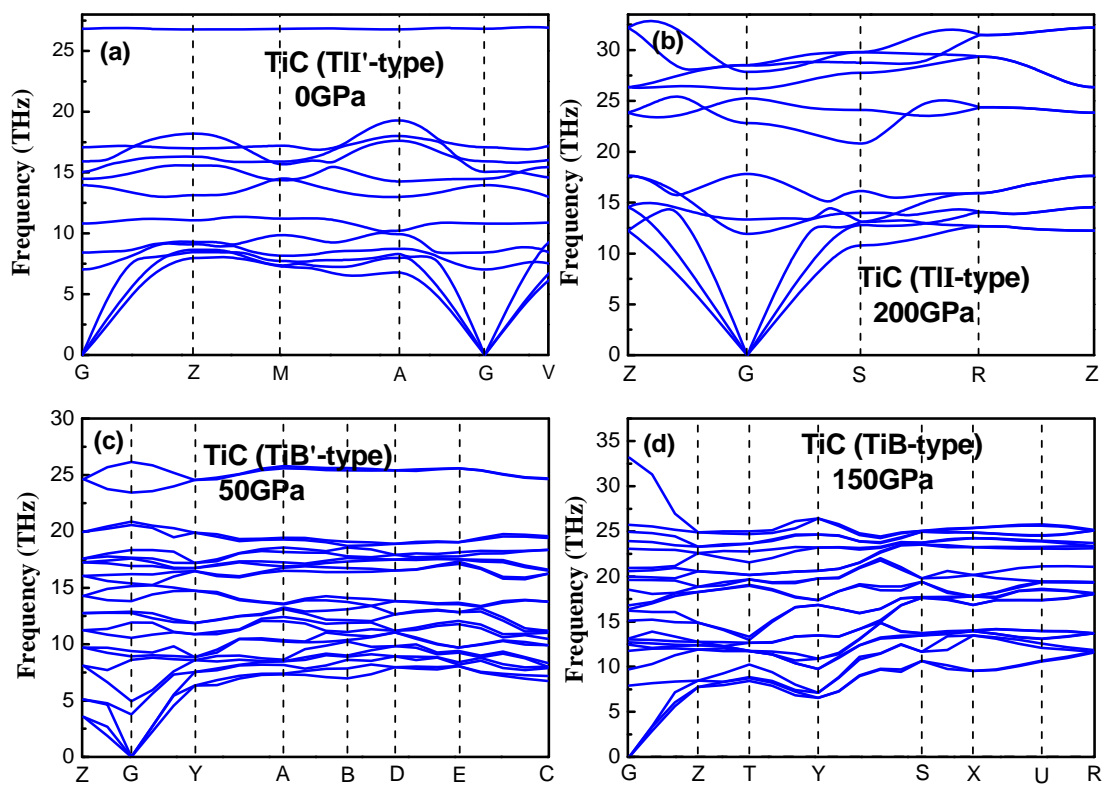

**Figure S1.** The calculated phonon dispersion curves for (a) the TiC (TlI′-type) phase at 0 GPa, (b) the TiC (TlI-type) phase at 200 GPa, (c) the TiC (TiB′-type) phase at 50 GPa, and (d) the TiC (TiB-type) phase at 150 GPa.

**Table S1.** The calculated elastic constants, bulk moduli $B$ and shear moduli $G$ of high-pressure phases of TMCs at the ground state.

| Phase | $C_{11}$ | $C_{22}$ | $C_{33}$ | $C_{44}$ | $C_{55}$ | $C_{66}$ | $C_{12}$ | $C_{13}$ | $C_{15}$ | $C_{23}$ | $C_{25}$ | $C_{35}$ | $C_{46}$ | B | G |
|---|---|---|---|---|---|---|---|---|---|---|---|---|---|---|---|
| TlI′-type | 510.9 | 433.1 | 415.7 | 140.7 | 181.5 | 162.6 | 103.2 | 130.1 | -47.8 | 111.7 | 26.0 | -34.8 | -4.7 | 227.8 | 164.6 |
| TiB′-type | 423.6 | 364.9 | 505.8 | 145.6 | 174.1 | 126.2 | 95.2 | 137.0 | 21.3 | 163.1 | -0.1 | -0.2 | 25.7 | 231.6 | 149.1 |
| TlI′-type | 336.7 | 357.8 | 286.5 | 58.7 | 139.9 | 129.7 | 86.4 | 120.1 | -36.3 | 71.4 | 37.2 | -5.1 | 14.2 | 170.7 | 112.5 |
| TlI′-type | 385.5 | 403.9 | 364.6 | 102.9 | 156.8 | 132.9 | 94.8 | 146.4 | -36.4 | 125.9 | 30.7 | -17.0 | 5.5 | 209.8 | 131.0 |
| TlI′-type | 557.1 | 420.1 | 369.6 | 75.3 | 152.9 | 144.2 | 124.0 | 195.1 | -106.6 | 133.6 | 49.0 | -13.9 | -34.6 | 250.3 | 134.1 |
| TiB′-type | 347.4 | 275.9 | 585.5 | 85.3 | 160.6 | 66.8 | 180.1 | 121.7 | 28.9 | 202.3 | -0.2 | -1.2 | 21.2 | 246.4 | 109.5 |
| TlI′-type | 386.3 | 383.4 | 331.9 | 69.1 | 129.0 | 112.2 | 173.7 | 256.4 | -89.7 | 138.3 | 48.1 | 14.8 | -20.7 | 248.7 | 97.6 |
| TlI′-type | 387.4 | 435.8 | 409.2 | 102.2 | 160.3 | 23.0 | 278.2 | 364.0 | -116.6 | 208.9 | 34.1 | 6.0 | -7.5 | 326.1 | 82.5 |

**Table S2.** The calculated bulk moduli and shear moduli for six *B*1 structured TMCs at ambient pressure, compared with the available experimental data.

| TMCs | *B* | *G* | TMCs | *B* | *G* |
|---|---|---|---|---|---|
| TiC | 242.1 | 157.8 | VC | 286.4 | 221.8 |
|  | 235±2[a] |  |  | 258 ± 11[b] |  |
| ZrC | 209.5 | 158.6 | NbC | 294.8 | 197.5 |
|  |  |  |  | 274 ± 3[c] |  |
| HfC | 242.2 | 178.2 | TaC | 309.7 | 224.6 |
|  |  |  |  | 345 ±9[d] |  |

[a] Dubrovinskaia, N. A.; Dubrovinsky, L. S.; Saxena, S. K.; Ahuja, R.; Johansson, B. *J. Alloys Compd.* **1999**, *289*, 24-27.

[b] Liermann, H. P.; Singh, A. K.; Manoun, B.; Saxena, S. K.; Prakapenka, V. B.; Shen, G. *Int. J. Refract. Met. Hard Mater.* **2004**, *22*, 129-132.

[c] Liermann, H. P.; Singh, A. K.; Somayazulu, M.; Saxena, S. K. *Int. J. Refract. Met. Hard Mater.* **2007**, *25*, 386-391.

[d] Liermann, H. P.; Singh, A. K.; Manoun, B.; Saxena, S. K.; Zha, C. S. *Int. J. Refract. Met. Hard Mater.* **2005**, *23*, 109-114.